\documentclass[aps,prb,twocolumn,superscriptaddress,showpacs,amsmath,amssymb,footinbib,longbibliography]{revtex4-2}
\usepackage[english]{babel}
\usepackage{graphicx}% Include figure files
\usepackage{dcolumn}% Align table columns on decimal point
\usepackage{bm}% bold math
\usepackage{subfigure}
\usepackage[utf8x]{inputenc}
\usepackage{color}
\usepackage{textcomp}
\usepackage[colorlinks,bookmarks=false,citecolor=blue,linkcolor=red,urlcolor=blue]{hyperref}
\newcommand{\op}[1]{%
    \fontdimen12\textfont3=2pt\fontdimen12\scriptfont3=1.4pt%
    \!\null\mathop{\vphantom{#1}\smash{#1}}\limits_{\sim}\null\!}
\newcommand{\xref}[1]{\protect\ref{#1}}
\newcommand{\figref}[1]{Fig.~\protect\ref{#1}}

\renewcommand{\eqref}[1]{Eq.~(\protect\ref{#1})}

\newcommand{\vecops}[1]{\op{\vec{s}}_{#1}}

\newcommand{\JS}[1]{#1}

\begin{document}

% %
\title{Theoretical investigations of tetrameric magnetic molecules for sub-kelvin cooling}
% %
\author{Dennis Westerbeck}
\author{J\"urgen Schnack}
\email{jschnack@uni-bielefeld.de}
\affiliation{Fakult\"at f\"ur Physik, Universit\"at Bielefeld, Postfach 100131, D-33501 Bielefeld, Germany}

\begin{abstract}
Magnetic molecules are a class of compounds that is also investigated 
in view of their magnetocaloric properties. The isothermal entropy change 
and the adiabatic temperature change are key figures of merit for 
magnetocaloric performance.
Here, we investigate spin systems of realistic molecular structures comprising 
four spins. In view of 
\JS{potentially} large spin quantum numbers as for gadolinium
we model these spin systems by a combination of Heisenberg and dipolar 
spin-spin interactions.
It turns out that a tetrahedral structure with ferromagnetic exchange
interactions yields the best figures of merit.
\end{abstract}

\maketitle

%%%%%%%%%%%%%%%%%%%%%%%%%%%%%%%%%%%%%%%%%%%%%%%%%%%%%%%%%%%%%%%%%%%%%%%%
\section{Introduction}
\label{sec-1}

One could summarize the efforts in the field of molecule-based 
magnetocalorics as looking for
``recipes for enhanced molecular cooling" \cite{EvB:DT10}.
Among the more general recipes is the suggestion to employ
frustrated spin systems \cite{Sch:DT10}, i.e., systems with competing exchange
interactions as for instance given in cuboctahedra or other
geometrically frustrated arrangements of spins
\cite{Zhi:PRB03,HoW:PB06,HoZ:JPCS09,SCM:NC14,BML:npjQM18}.
Here, the idea is to generate an enhanced low-lying density of states
that can be manipulated by moderate magnetic fields. 
If the system is close to a quantum critical point this effect is
particularly large \cite{ZGR:PRL03,WTJ:PNAS11,BML:npjQM18,RSH:ZNA24}.
On the other hand, spin systems with a large energy gap between 
the ground state and excited states 
would not be as well suited.

Such an energy gap occurs for instance in spin systems termed single-molecule magnets
that are characterized by a strong easy-axis anisotropy that separates a 
two-fold degenerate ground state from higher-lying levels.
However, such spin systems show a rotational magnetocaloric effect
that occurs when the magnitude of the magnetic field is kept constant
and the sample (or the field) are rotated 
\cite{LRE:ANIE16,OGG:SR18,TTO:PB18,BES:JMMM19}. Although very appealing, technical
difficulties such as heat from friction prevent a use at low temperatures so far.

Another trend in the field is to employ lanthanides, in particular gadolinium, 
because of the large spin and corresponding large number of levels per spin that directly 
translate into available entropy \cite{SZT:CC11,HSP:ACIE12,PZK:JACS12,TGD:JACS23,ZCE:A24}.
Larger spins and in particular net spins per molecules might, however,
lead to long range magnetic order that precludes the desired low-temperature
(sub-kelvin) cooling \cite{ELJ:JMC06}. Very low temperatures are thus achieved
when separating the molecules or spins sufficiently well \cite{MME:AM12,MDG:JACS25}.

In this article, we are going to investigate spin systems composed of four
spins with realistic geometries as obtained in chemical synthesis. This investigation
is along the ideas of Refs.~\cite{GCS:APL13,ELP:APL14,GCS:APL14} where square-pyramidal
structures have been studied employing the Heisenberg model. 
In our current study, 
we add intramolecular dipolar interaction which can be sizable for larger spins \cite{PLZ:CS16}.
It also plays an important role below about one kelvin.
\JS{For reasons of computational feasibility we study only single-spin quantum numbers of 
$s=3/2$ in this article. The qualitative conclusions are not altered by this restriction.}

Our investigations suggest that tetrahedral arrangements of spins with 
ferromagnetic exchange interactions yield the strongest magnetocaloric effect. 
Furthermore, among the investigated structures it is the only arrangement 
that is able to achieve very low temperatures.
We will substantiate our findings with a comparison of four typical 
\JS{geometrical}
structures, however, limit ourselves to a few central graphs. The reader is encouraged 
to have a look at a recent Ph.D. thesis \JS{(in German)} offering detailed explorations 
of the possible huge parameter space \cite{Westerbeck:Diss25}. 

The paper is organized as follows. In Section \ref{sec-2} we introduce 
the model and major observables before we present our results in Sec.~\xref{sec-3}.
The article closes with a discussion in Section~\ref{sec-4}.

%%%%%%%%%%%%%%%%%%%%%%%%%%%%%%%%%%%%%%%%%%%%%%%%%%%%%%%%%%%%%%%%%%%%%%%%
\section{Model}
\label{sec-2}

In the present study we employ a Hamiltonian that consists of 
Heisenberg exchange interactions as well as dipolar interactions,
%--------------------------------------------------------
\begin{align}\label{eq:mce-hamop}
	\op{H}=&\sum_{i<j} J_{ij} \vecops{i}\cdot\vecops{j}+ g\mu_B  B_z \sum_i \op{s}_i^z
	\\
	&+ \frac{\mu_0\mu_B^2}{4\pi}\sum_{i<j}\frac{g^2}{r_{ij}^3}
	\left(\op{\vec{s}}_i\cdot \op{\vec{s}}_j-3\cdot\op{\vec{s}}_i\cdot 
	\vec{e}_{ij}\otimes\vec{e}_{ij}\cdot \op{\vec{s}}_j\right)
\nonumber
    \ .
\end{align}
%--------------------------------------------------------
Here, $\op{\vec{s}}_i$ denotes the spin vector operator at site $i$, and 
a tilde is used to denote operators in general.
The Zeeman term is assumed along the global $z$-direction,
thus the molecules have to be rotated to investigate the influence
of the relative orientation which, however, is expected to be small 
since the anisotropic effect of the dipolar interaction is small.
Therefore, we will show only one chosen direction for each example. 
In addition, we fix $g=2$ in the following.
Single-ion anisotropy is not considered in this study although of course 
potentially also important, compare e.g.\ \cite{GZS:PRB05,SWB:IC23}.
\JS{The collected experience of the field is that the zero-field splitting induced by
single-ion anisotropy precludes cooling to very low temperatures
due to the generation of larger excitation gaps and thus
a reduction of the low-lying density of states \cite{EvB:DT10}.}

\JS{The spin systems are treated by numerically diagonalizing the Hamiltonian matrix
for every investigated pair of exchange interactions, every external field and every
dipolar arrangement. Thermal equilibrium 
observables are calculated in the canonical ensemble using the eigenvalues and the
eigenvectors of the Hamiltonian matrix.} We concentrate on 
realistic exchange interactions and a
realistic field change from $B=7$~T to $B=0$ when evaluating the isothermal
entropy change.

\begin{table}
\caption{\label{tetracaloric-t-1}
Investigated structures and considered exchange interactions. $J_1$ is shown in red, $J_2$ in blue.}
\begin{tabular}{c} 
\includegraphics[width=0.2\textwidth]{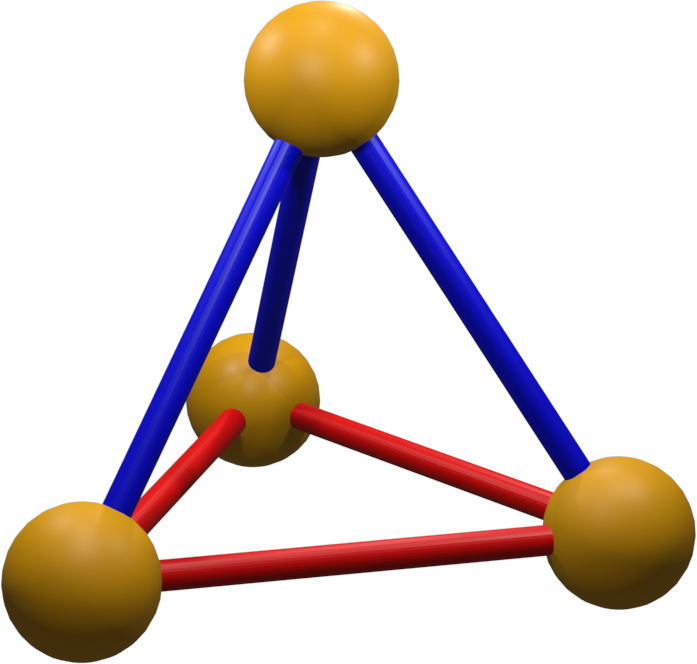}\\
\phantom{x}\hfill $J_1=J_{\text{base}}=\{J_{12}, J_{23}, J_{13} \}$, $J_2=J_{\text{top}}=\{J_{14}, J_{24}, J_{34} \}$ \hfill\phantom{x} \\[1mm] 
\hline\\[1mm] 
\includegraphics[width=0.2\textwidth]{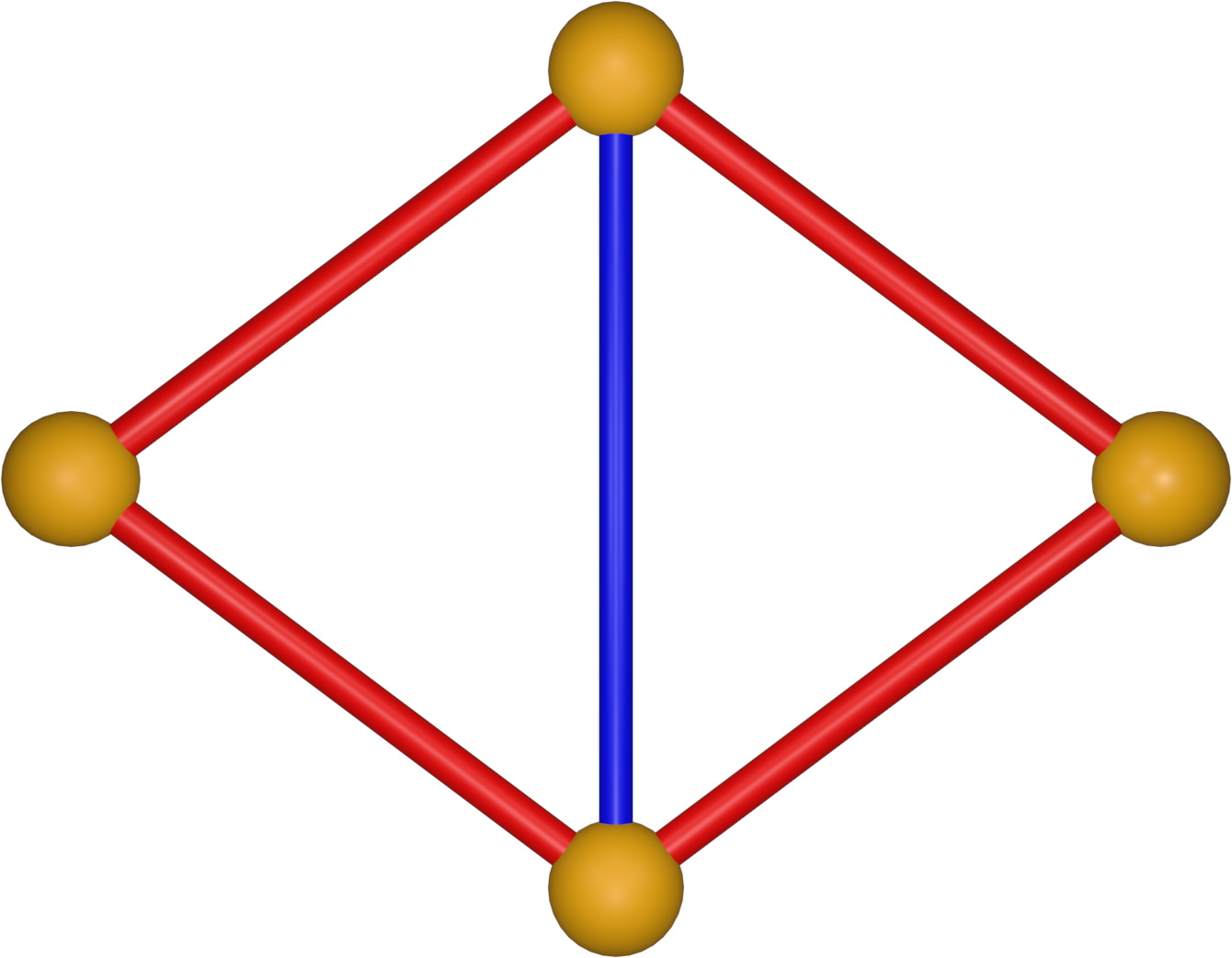}\\
\phantom{x}\hfill $J_1=J_{\text{wings}}=\{J_{12}, J_{23}, J_{34}, J_{14} \}$, $J_2=J_{\text{body}}=\{J_{13} \}$ \hfill\phantom{x} \\[1mm] 
\hline\\[1mm] 
\includegraphics[width=0.2\textwidth]{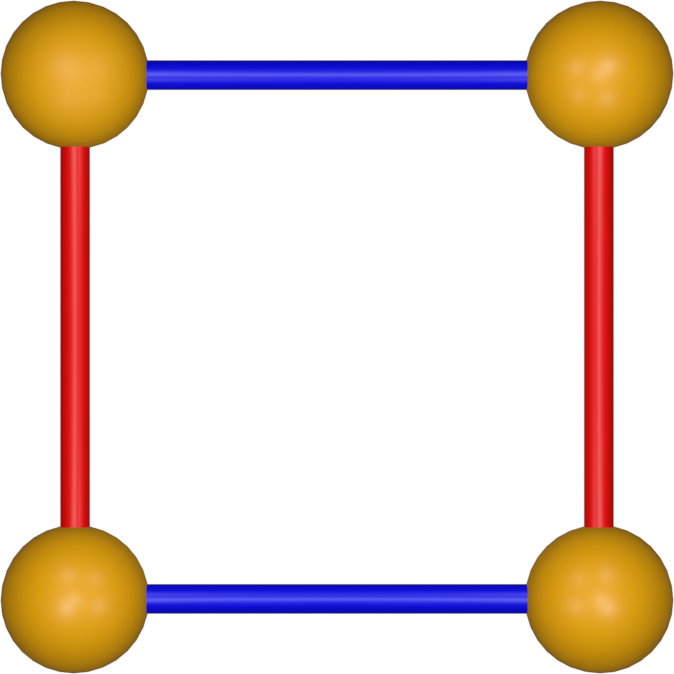}\\
\phantom{x}\hfill $J_1=J_{\text{outer}}=\{J_{12}, J_{34} \}$, $J_2=J_{\text{inner}}=\{J_{23} \}$ \hfill\phantom{x} \\[1mm] 
\hline\\[1mm] 
\includegraphics[width=0.2\textwidth]{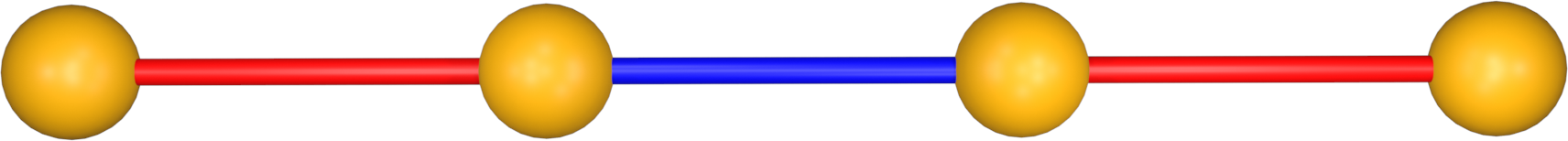}\\
\phantom{x}\hfill $J_1=J_{\text{ver}}=\{J_{12}, J_{34} \}$, $J_2=J_{\text{hor}}=\{J_{23}, J_{14} \}$ \hfill\phantom{x} \\[1mm] 
\hline
\end{tabular}
\end{table}

%%%%%%%%%%%%%%%%%%%%%%%%%%%%%%%%%%%%%%%%%%%%%%%%%%%%%%%%%%%%%%%%%%%%%%%%
\section{Results}
\label{sec-3}

In the following, we consider four different typical spin systems as listed in table~\xref{tetracaloric-t-1},
a tetrahedron, a butterfly system, a linear chain, and a square. We allow two different exchange
interactions in our search for advantageous coupling schemes.
We show isothermal entropy changes in the parameter 
space of $J_1$ and $J_2$ for the spin quantum number $s=3/2$
for a field sweep from $B=7$~T down to $B=0$.
The Heisenberg limit is denoted as $d=\infty$, i.e., for zero dipolar interaction.
In order to account for dipolar influence we pick some reasonable 
(but not too large) distances. It is clear that the parameter space of such
an investigation is practically limitless since one could also vary the spin 
quantum numbers, introduce more exchange constants, and vary the distances 
that are important for the dipolar interaction in a much larger range.
Here we focus on central and typical results, see also \cite{Westerbeck:Diss25}.

%===================    figure   =================================
\begin{figure}[ht!]
\centering
\includegraphics*[width=0.85\columnwidth]{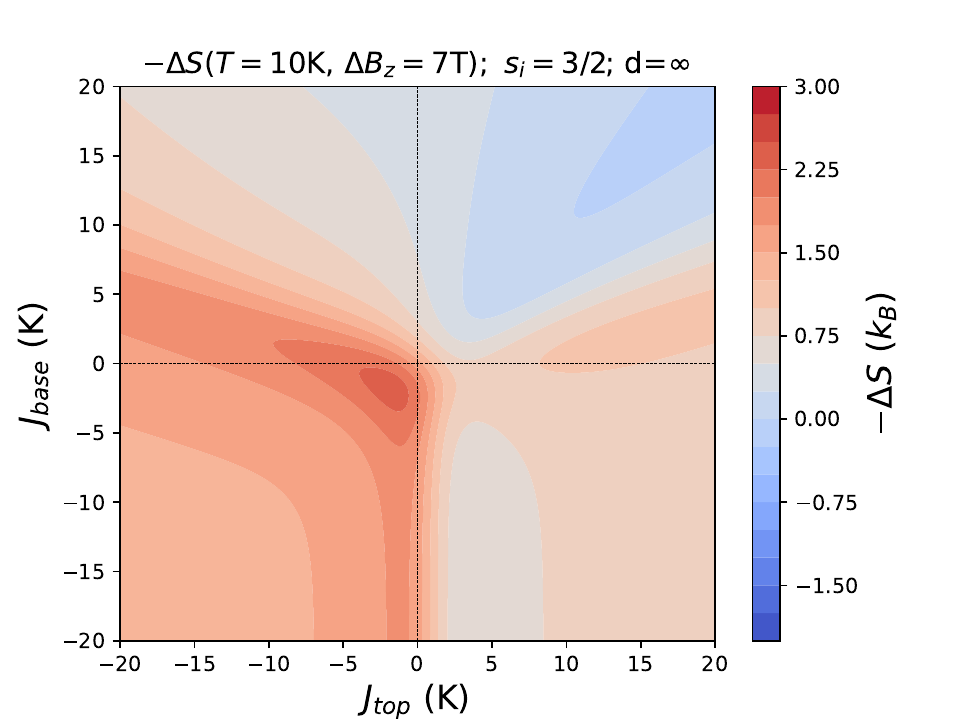}

\includegraphics*[width=0.85\columnwidth]{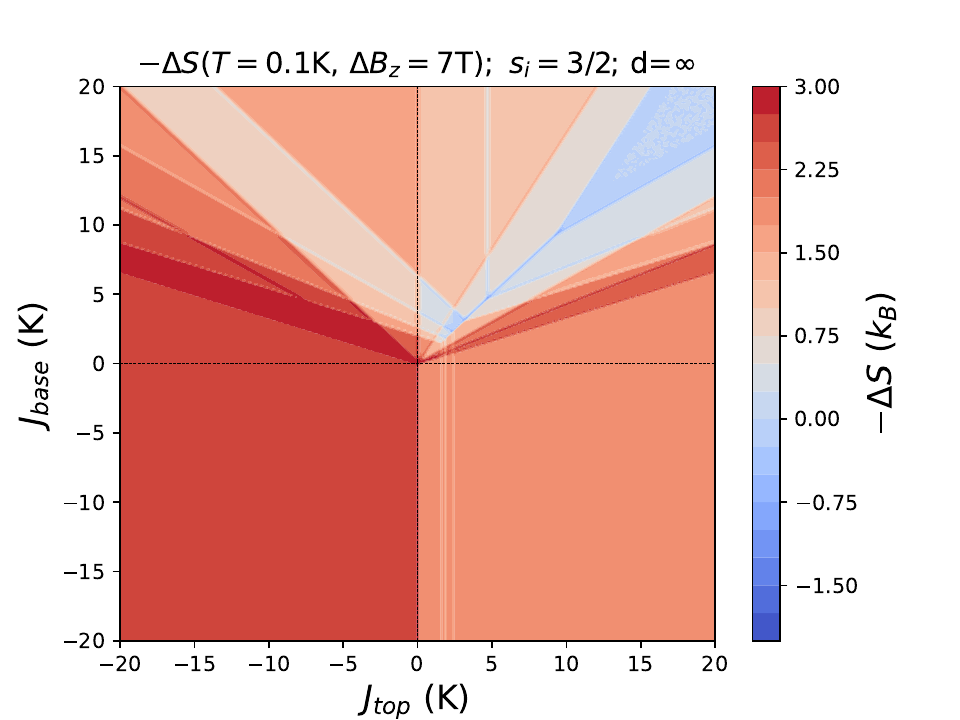}

\includegraphics*[width=0.85\columnwidth]{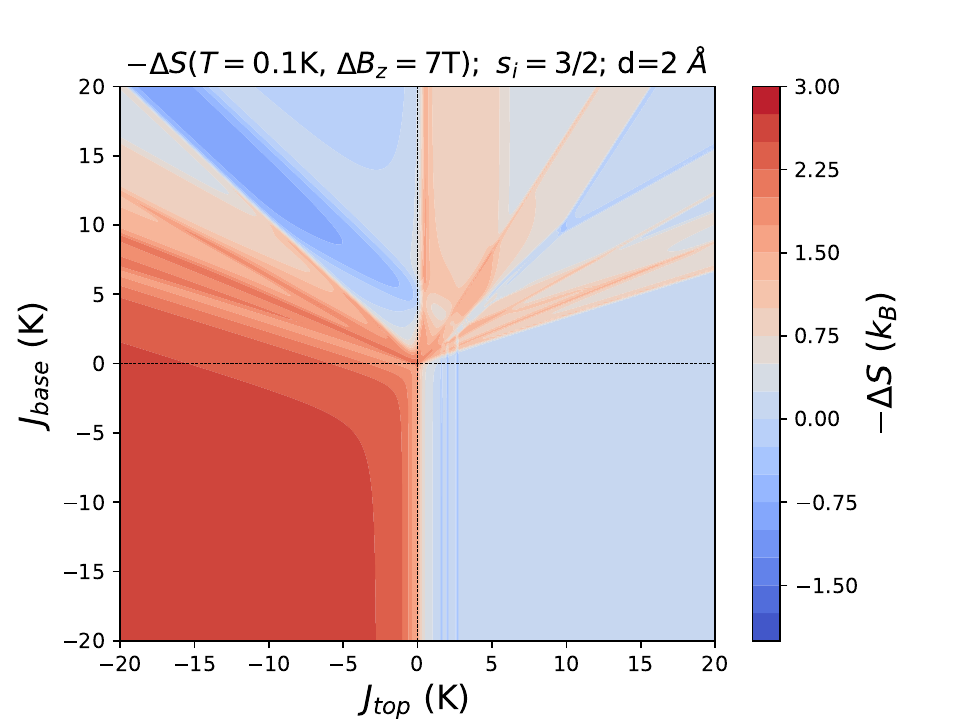}
\caption{Isothermal entropy change for a tetrahedron with $s=3/2$ at $T=10$~K (top), 
$T=0.1$~K (middle), and $T=0.1$~K (bottom);
the latter including dipolar interactions for a distance between of $d=2$~\AA.
The magnetic field is perpendicular to the plane of the red triangle in the tetrahedron, 
see Tab.~\ref{tetracaloric-t-1}.
The color maps show the respective entropy changes.}
\label{tetracaloric-f-1}
\end{figure}
%===================    figure   =================================

For the discussion of the isothermal entropy change we picked two temperatures,
$T=10$~K and $T=0.1$~K. 
The behavior at $T=0.1$~K without dipolar interactions is then contrasted
with the behavior including dipolar interactions. At $T=10$~K the dipolar interaction 
has practically no effect.

%===================    figure   =================================
\begin{figure}[ht!]
\centering
\includegraphics*[width=0.85\columnwidth]{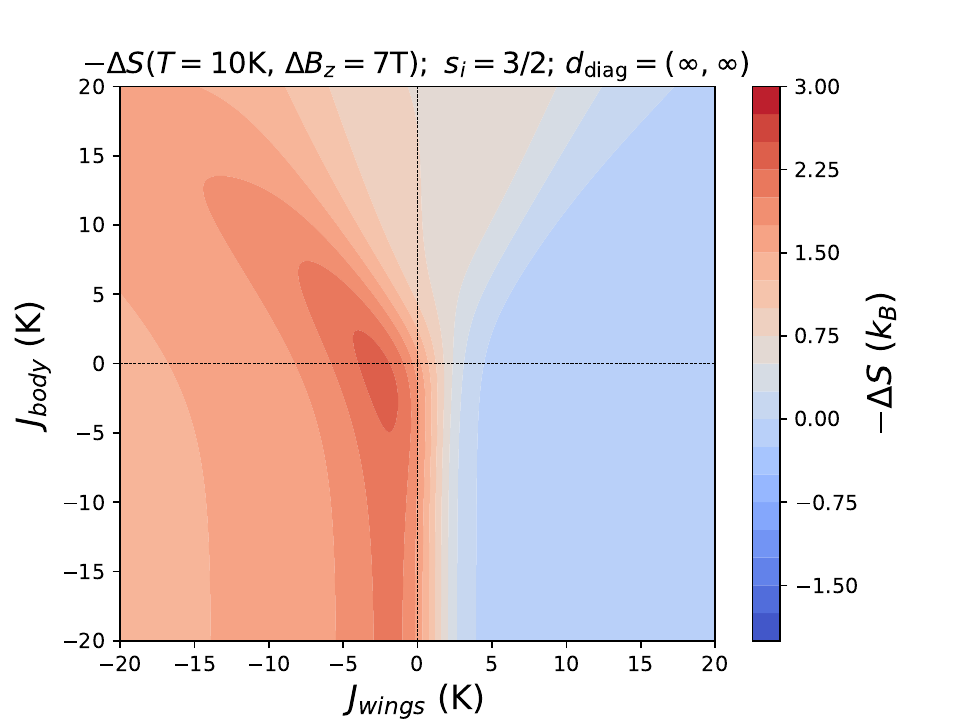}

\includegraphics*[width=0.85\columnwidth]{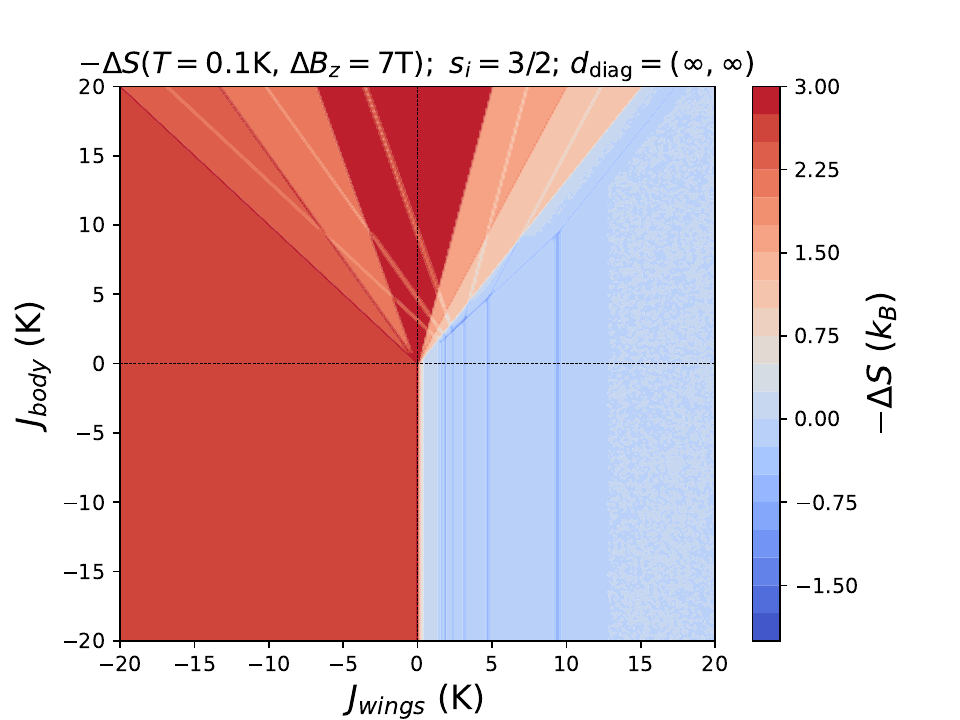}

\includegraphics*[width=0.85\columnwidth]{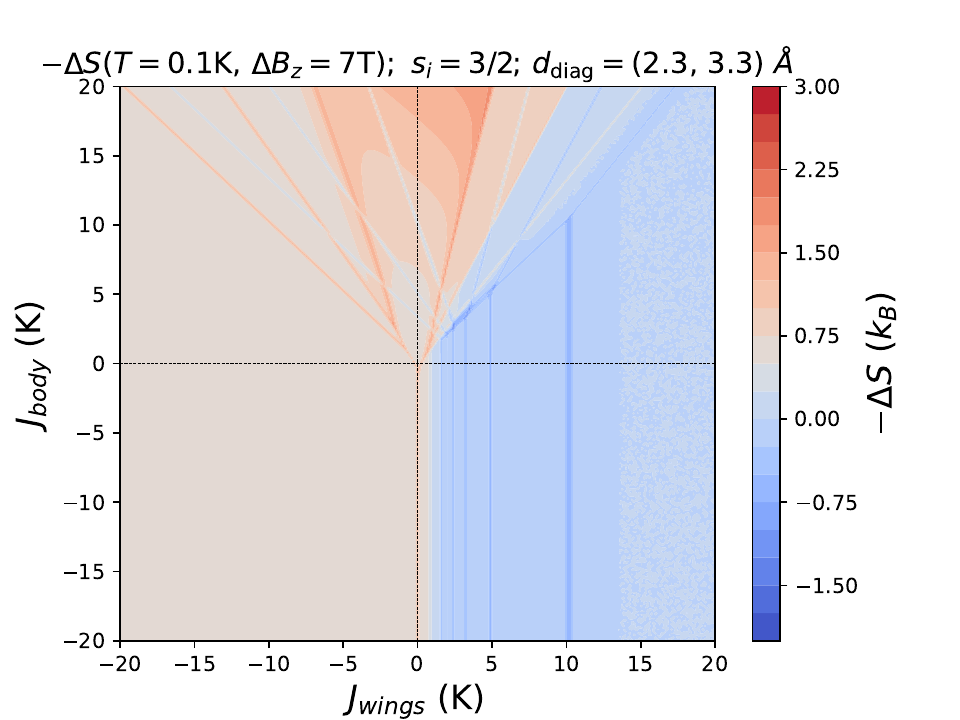}
\caption{Isothermal entropy change for a butterfly with $s=3/2$ at $T=10$~K (top), 
$T=0.1$~K (middle), and $T=0.1$~K (bottom);
the latter including dipolar interactions for distances of $d_1=2.3$~\AA\ 
(corresponding to the exchange path $J_{13}$)
and $d_2=3.3$~\AA\ (corresponding to the exchange path $J_{24}$).
The magnetic field points along the longer diagonal of the butterfly 
corresponding to the exchange path $J_{24}$, 
see Tab.~\ref{tetracaloric-t-1}.
The color maps show the respective entropy changes.}
\label{tetracaloric-f-2}
\end{figure}
%===================    figure   =================================

At $T=0.1$~K the dipolar interaction has got an important effect. Without
dipolar interaction the largest isothermal entropy change is typically 
achieved for non-interacting spins. This is unrealistic since at small enough 
temperatures the effect of dipolar interactions will not only be visible,
it will be unavoidable. In molecular
magnetism this typically happens below 1~K. Therefore, when discussing sub-kelvin 
cooling the dipolar interaction has to be taken into account.

%===================    figure   =================================
\begin{figure}[ht!]
\centering
\includegraphics*[width=0.85\columnwidth]{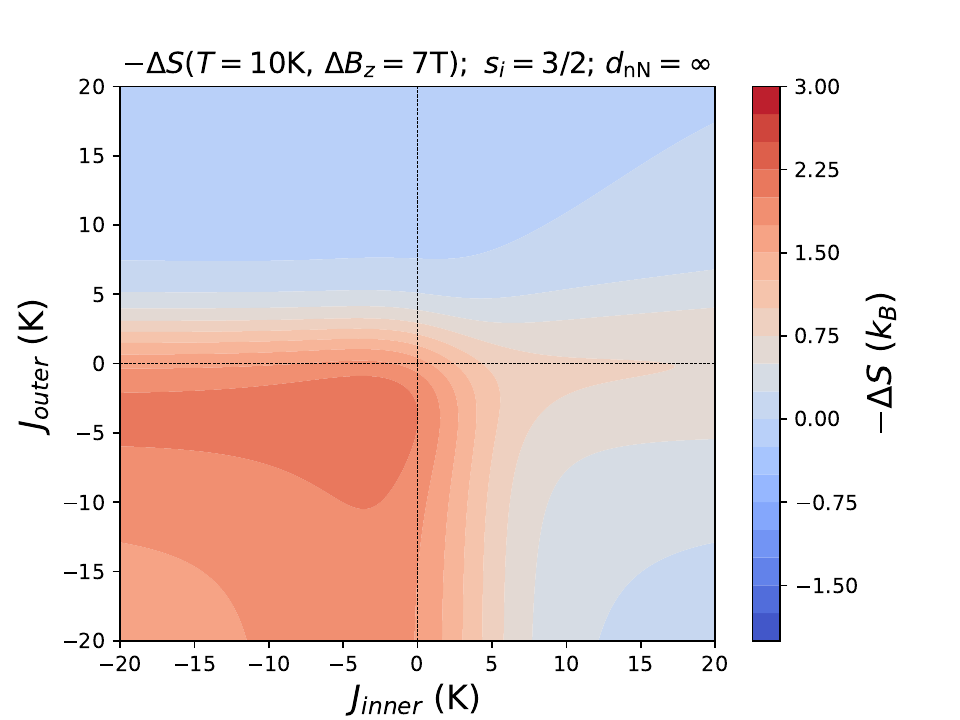}

\includegraphics*[width=0.85\columnwidth]{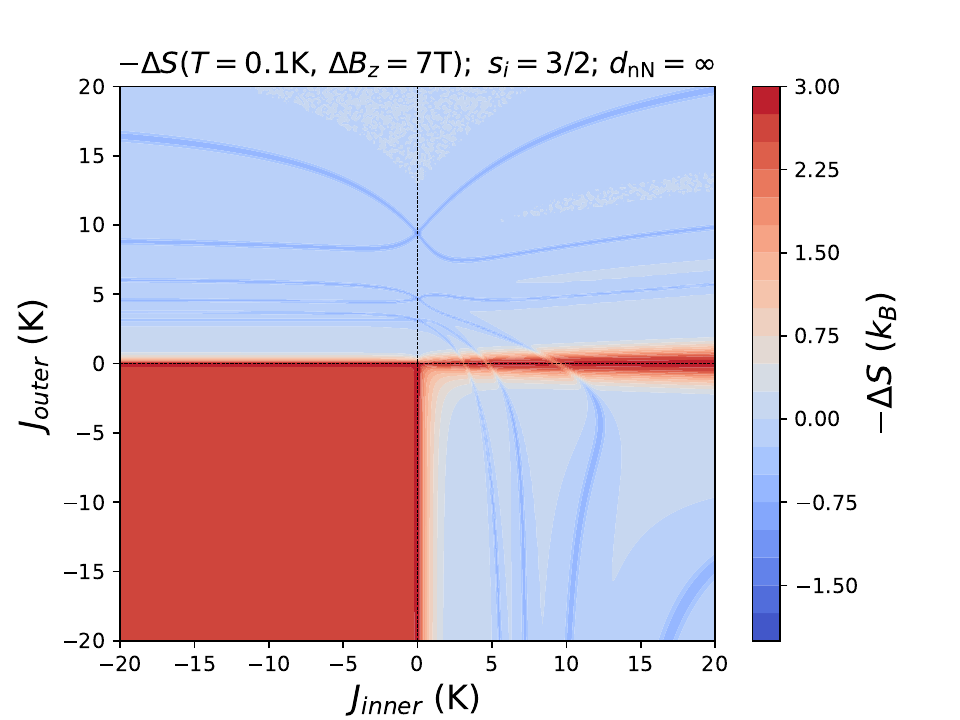}

\includegraphics*[width=0.85\columnwidth]{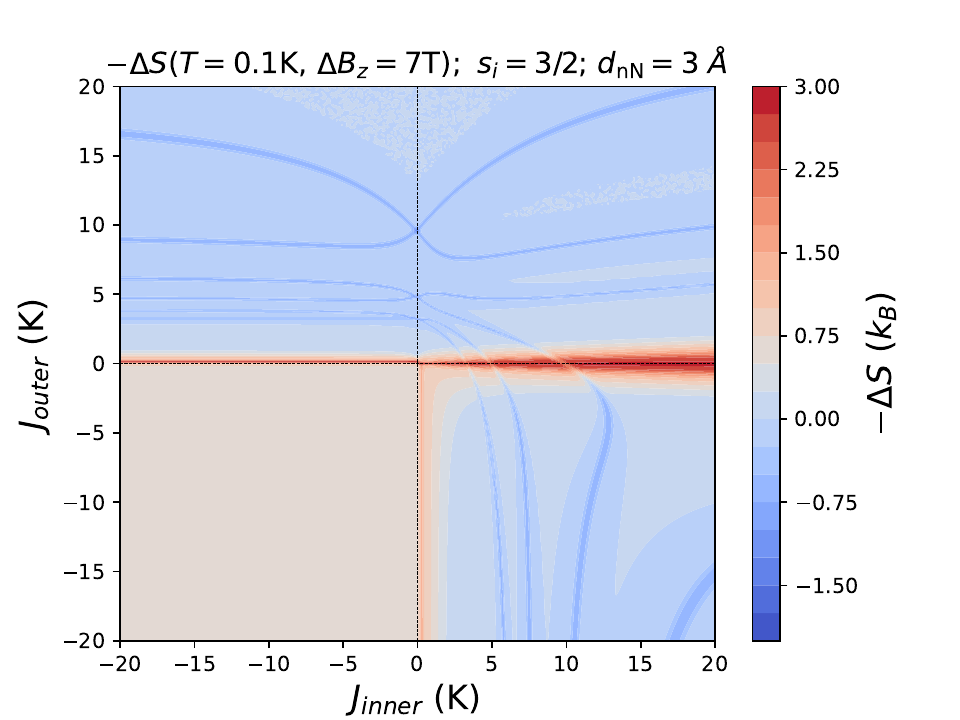}
\caption{Isothermal entropy change for a chain with $s=3/2$ at $T=10$~K (top), 
$T=0.1$~K (middle), and $T=0.1$~K (bottom);
the latter including dipolar interactions for distances 
between adjacent spins of $d=3$~\AA.
The magnetic field points along the axis of the chain.
The color maps show the respective entropy changes.}
\label{tetracaloric-f-3}
\end{figure}
%===================    figure   =================================

%===================    figure   =================================
\begin{figure}[ht!]
\centering
\includegraphics*[width=0.85\columnwidth]{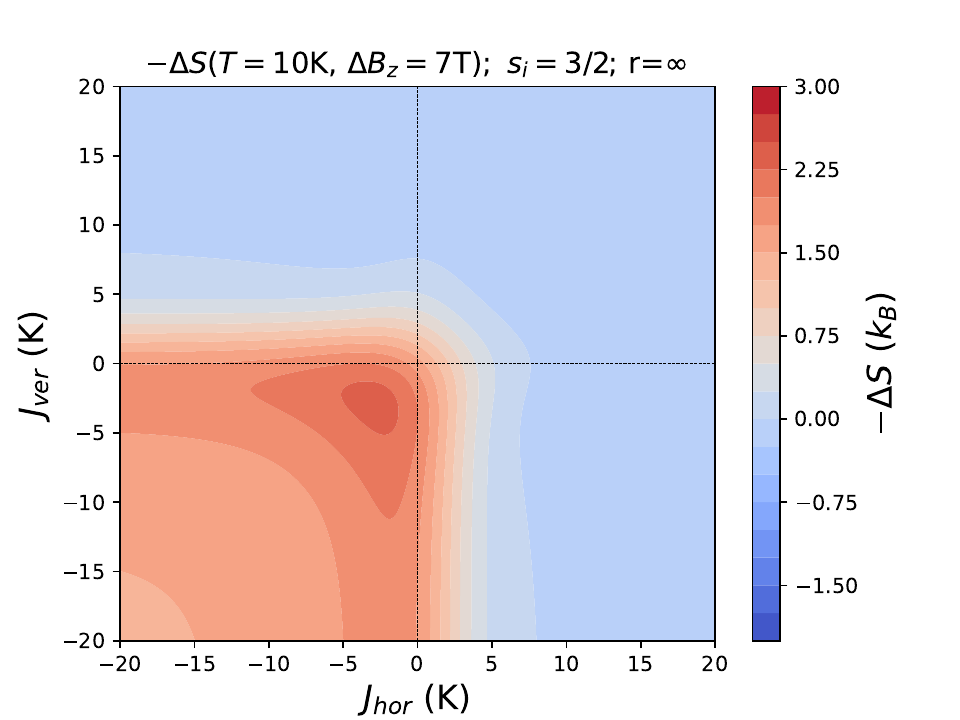}

\includegraphics*[width=0.85\columnwidth]{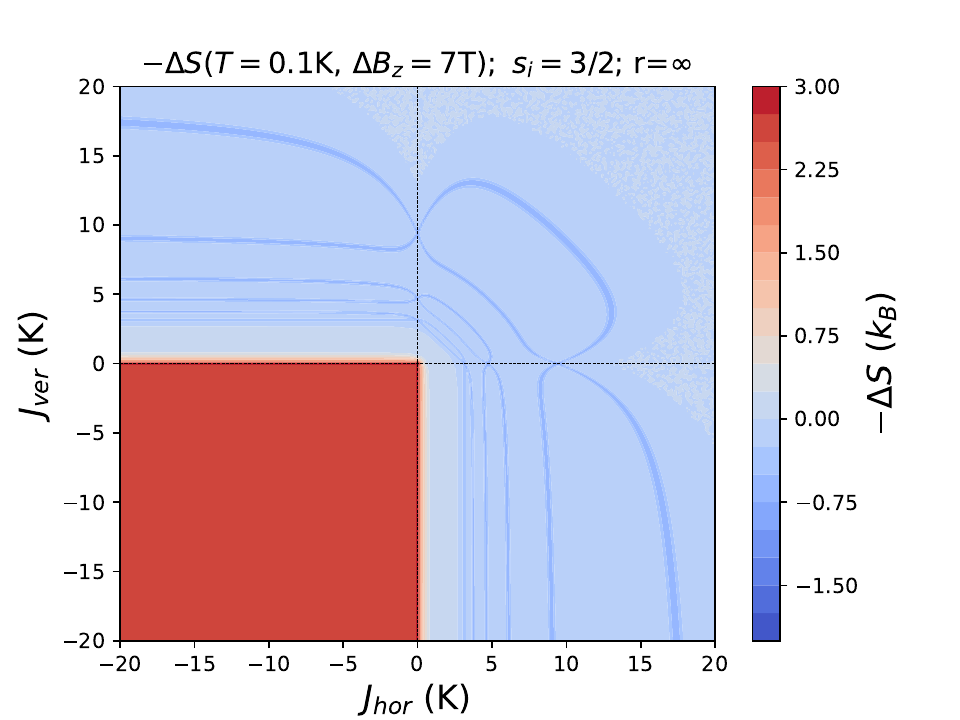}

\includegraphics*[width=0.85\columnwidth]{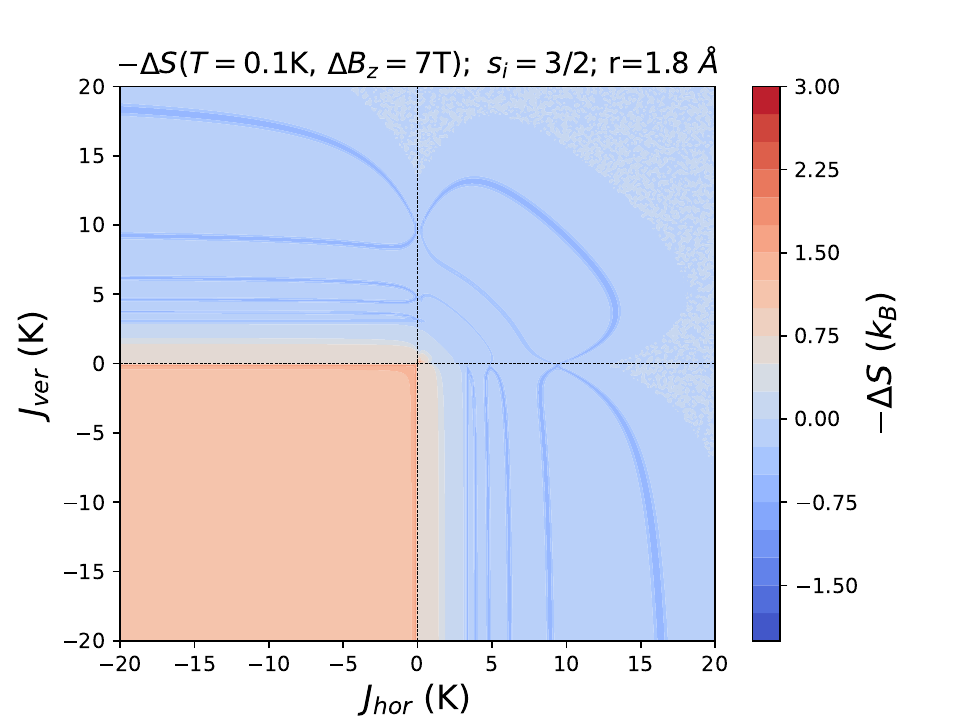}
\caption{Isothermal entropy change for a square with $s=3/2$ at $T=10$~K (top), 
$T=0.1$~K (middle), and $T=0.1$~K (bottom);
the latter including dipolar interactions for distances of $d=2.54$~\AA\
between adjacent spins.
The magnetic field points along two parallel edges of the square.
The color maps show the respective entropy changes.}
\label{tetracaloric-f-4}
\end{figure}
%===================    figure   =================================

Figure~\xref{tetracaloric-f-1} shows the isothermal entropy change for a tetrahedron 
with spin quantum number $s=3/2$. The upper and middle panel are for $T=10$~K and $T=0.1$~K,
respectively, both without dipolar interactions. The lower panel displays the situation
at $T=0.1$~K with dipolar interaction. The most reddish parts denote the largest
isothermal entropy changes. For $T=10$~K, where the dipolar interaction does not play
much of a role, the optimal molecule would possess very small and slightly ferromagnetic
couplings. For $T=0.1$~K, one notices that the very large isothermal entropy change
in the quadrant of both ferromagnetic exchange interactions is not much altered 
by dipolar interactions. This robust behavior singles out the tetrahedron among the 
discussed four-spin structures.

The butterfly and the chain, Figs.~\xref{tetracaloric-f-2} and \xref{tetracaloric-f-3},
lose their good isothermal entropy changes for ferromagnetic couplings 
at sub-kelvin temperatures when the dipolar interaction is taken into account.
This effect is rather drastic as can be seen in the bottom panels of both figures.
For $T=10$~K, the chain shows good isothermal entropy changes for small 
ferromagnetic interactions whereas the butterfly performs good for
ferromagnetic $J_{\text{wings}}$ independent of the sign of $J_{\text{body}}$,
compare top panel of \figref{tetracaloric-f-2}. In view of existing chemical
compound this is a clear advantage.

The square, \figref{tetracaloric-f-4}, performs best for ferromagnetic exchange 
interactions. When dipolar interactions are considered the isothermal entropy change
is not as bad as for the butterfly and the chain, but not as good as for the tetrahedron.

%%%%%%%%%%%%%%%%%%%%%%%%%%%%%%%%%%%%%%%%%%%%%%%%%%%%%%%%%%%%%%%%%%%%%%%%
\section{Discussion}
\label{sec-4}

In this paper, we investigated four typical geometric structures of four
spins each as they often appear as motives in magnetic molecules.
It turns out that a tetrahedral arrangement with ferromagnetic 
exchange interactions is most promising in view 
of the isothermal entropy change that is robust under the influence 
of dipolar interactions at sub-kelvin temperatures. Even combinations 
of ferromagnetic and antiferromagnetic interactions yield a reasonable 
magnetocaloric performance for the tetrahedron. 

%===================    figure   =================================
\begin{figure}[ht!]
\centering
\includegraphics*[width=0.85\columnwidth]{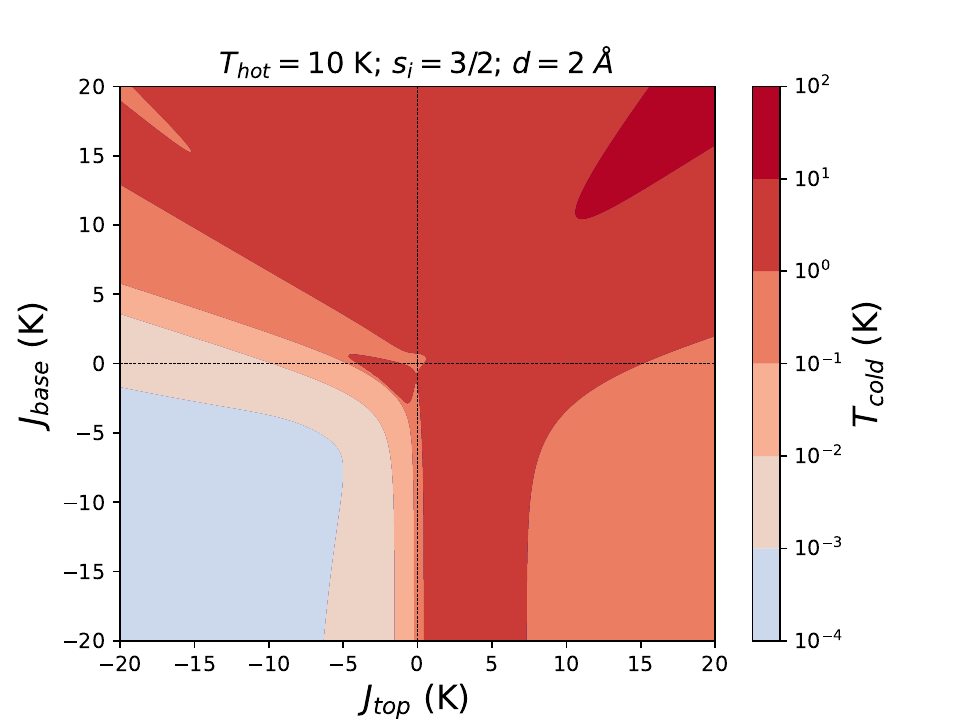}
\caption{Adiabatic temperature change for a tetrahedron with $s=3/2$ with 
$B_{\text{hot}}=7$~T, $B_{\text{cold}}=0$, and $T_{\text{hot}}=10$~K.
Dipolar interactions for distances of $d=2$~\AA\ are included.
The magnetic field is perpendicular to the plane of the red triangle 
in the tetrahedron, compare Tab.~\ref{tetracaloric-t-1}.
The color map shows the respective $T_{\text{cold}}=0$ 
on a logarithmic scale.}
\label{tetracaloric-f-5}
\end{figure}
%===================    figure   =================================

The adiabatic temperature change is shown for a tetrahedron in \figref{tetracaloric-f-5}.
One notices that low temperatures can be reached for large parts of the phase
diagram but not for antiferromagnetic interactions. 
However, strong ferromagnetic interactions lead to the best result
and push the limits of magnetocaloric refrigeration to very low temperatures
(even milikelvin for the investigated scenarios). 
This trend is largely independent of 
the specific $J_1$ and $J_2$ \cite{Westerbeck:Diss25}.

In our theoretical 
calculations low temperatures can be achieved even in regions where the isothermal 
entropy change would be small, e.g., in the lower right quadrant. For a true
periodic process such as a Carnot process one would of course prefer parameters
that yield a not too small heat transfer per cycle, see also \cite{ELP:APL14} 
on issues of realistic cycles. 

\begin{figure}[ht!]
\centering
\includegraphics*[width=0.85\columnwidth]{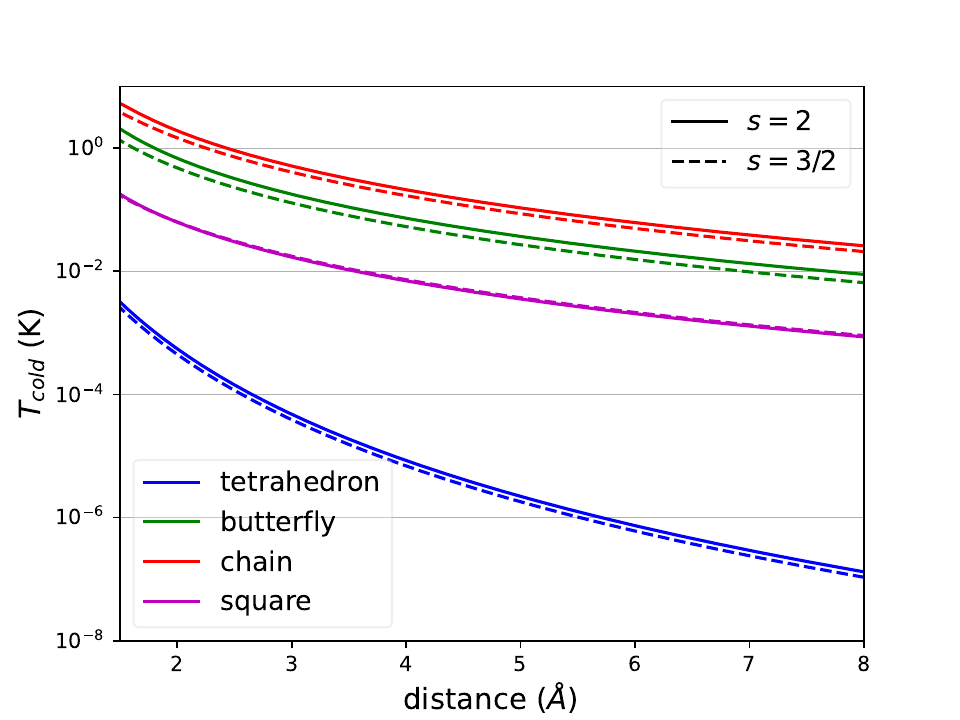}
\caption{Adiabatic temperature change for all four investigated structures
as function of the distance between spins for two spin quantum numbers;
$B_{\text{hot}}=7$~T, $B_{\text{cold}}=0$, and $T_{\text{hot}}=10$~K.
The exchange is ferromagnetic, all $J=-20$~K.}
\label{tetracaloric-f-6}
\end{figure}

Finally, \figref{tetracaloric-f-6} compares the four investigated tetra\-nuclear 
spin systems with ferromagnetic coupling ($J=-20$~K) in view of the adiabatic temperature change.

Again, $B_{\text{hot}}=7$~T, $B_{\text{cold}}=0$, and $T_{\text{hot}}=10$~K.
One sees that the tetrahedron outperforms all other three structures in terms
of the achievable $T_{\text{cold}}$ for every investigated distance of spins, i.e.,
for every strength of the dipolar interaction. This seems to be largely independent 
of spin quantum number. We speculate that in the 
case of the tetrahedron the dipolar interaction does not split the ground multiplet 
as much as it does for the other three structures. The magnitude of this splitting
sets the scale for the lowest achievable temperatures.

However and unfortunately, tetrahedra with larger spins -- iron(III) or gadolinium(III) --
seem to dominantly exhibit antiferromagnetic interactions, 
see e.g.\ \cite{PBH:CC94,MZG:NJC00,MZG:ANIE00,KMK:JACS06,RoU:DT06,KLZ:IC09}. 

%===================    figure   =================================
%===================    figure   =================================

Nickel-based tetrahedra may show ferromagnetic exchanges, see e.g.\ \cite{YWH:P03,MHR:EJIC04}
as well as some rare manganese compounds \cite{MPM:CC07} do, 
although for nickel also antiferromagnetic couplings are reported \cite{MBK:IC00}.
However, the larger problem with nickel and manganese is the 
non-negligible single-ion anisotropy 
and the possible biquadratic exchange for nickel that would very likely alter the
magnetocaloric behavior at sub-kelvin temperatures 
\JS{in an unfavorable way \cite{EvB:DT10}.}
 
Summarizing, we think that trying to synthesize ferromagnetic tetrahedra
is a worthwhile effort. This potentially includes tetrahedral chains 
\cite{SRV:PRE14,GaJ:PA17}.

\vspace*{3mm}

%%%%%%%%%%%%%%%%%%%%%%%%%%%%%%%%%%%%%%%%%%%%%%%%%%%%%%%%%%%%%%%%%%%%%%%%
\section*{Acknowledgment}

This work has received support from the EU via MSCA-DN MolCal,101119865. 
We thank Euan Brechin for fruitful discussions.

%%%%%%%%%%%%%%%%%%%%%%%%%%%%%%%%%%%%%%%%%%%%%%%%%%%%%%%%%%%%%%%%%%%%%%%%
%\bibliography{js-own.bib,js-other.bib}


\end{document}